\documentclass{ws-ijmpe}

\begin{document}

\markboth{D. Peressounko}{Bose-Einstein correlations of direct
photons in Au+Au collisions}

\catchline{}{}{}{}{}
%%%%%%%%%%%%%%%%%%%%%%%%%%%%%%%%%%%%%%%%%%%%%%%%%%%%%%%%%%%%%%%%%%%%

\title{Bose-Einstein correlations of direct
photons in Au+Au collisions at $\sqrt{s_{NN}} = 200$~GeV}

\author{D. Peressounko for the PHENIX collaboration\footnote{For the full list of the
PHENIX collaboration and acknowledgments, see$^9$.}}

\address{RRC "Kurchatov Institute", Kurchatov sq.1,\\
Moscow, 123182, Russia\\
peressou@rcf.rhic.bnl.gov}

\maketitle

\begin{history}
\received{(received date)}
\revised{(revised date)}
%\accepted{(Day Month Year)}
%\comby{(xxxxxxxxxx)}
\end{history}

\begin{abstract}
The current status of the analysis of direct photon Bose-Einstein
correlations in Au+Au collisions at $\sqrt{s_{NN}}=200$~GeV done
by the PHENIX collaboration is summarized. All possible sources of
distortion of the two-photon correlation function are discussed
and methods to control them in the PHENIX experiment are
presented.
\end{abstract}

\section{Introduction}

Photons have an extremely long mean free path length and escape
from the hot matter without rescattering. By measuring their
Bose-Einstein (or Hanbury-Brown Twiss, HBT) correlations one can
extract the space-time dimensions of the hottest central part of
the collision\cite{Makhlin,Srivastava,Peressounko,Alam-gg,Renk} in
contrast to hadron HBT correlations which measure the size of the
system at the moment of its freeze-out. Moreover, photons emitted
at different stages of the collision dominate in different ranges
of transverse momentum\cite{g-spectra}, therefore measuring photon
correlation radii at various average transverse momenta ($K_T$)
one can scan the space-time dimensions of the system at various
times and thus trace the evolution of the hot matter.

Photons emitted directly by the hot matter -- direct  photons --
constitute only a small fraction of the total photon yield while
the dominant contribution comes from decays of the final state
hadrons, mainly $\pi^0\to 2\gamma$ and $\eta\to 2\gamma$ mesons.
Fortunately, the lifetime of these hadrons is extremely large and
the width of the Bose-Einstein correlations between the decay photons is
of the order of a few eV and cannot obscure the direct photon
correlations. This feature can be used to extract the direct
photon yield\cite{Peressounko}: assuming that direct photons are
emitted incoherently, the photon correlation strength parameter
can be related to the proportion of direct photons as $\lambda =
1/2(N_\gamma^{dir}/N_\gamma^{incl})^2$. This approach is probably
the only way to experimentally measure direct photon yield at very
small $p_T$. Presently, the only experiment to have measured direct photon
Bose-Einstein correlations in ultrarelativistic heavy ion
collisions is WA98\cite{wa98}. An invariant correlation radius was
extracted and the direct photon yield was measured in Pb+Pb collisions
at $\sqrt{s_{NN}}=17$~GeV.

Since the strength of the direct photon Bose-Einstein correlation
is typically a few tenths of a percent, it is important to exclude all
background contributions which could distort the photon correlation
function. These contributions can be classified as following:
apparatus effects (close clusters interference -- "attraction" of
close clusters in the calorimeter during reconstruction) and
correlations caused by real particles. The latter in turn can be
divided into contribution due to "splitting" of particles --
processes like antineutron annihilation in the calorimeter and
photon conversion on detector material in front of the
calorimeter; contamination by correlated hadrons (e.g.
Bose-Einstein-correlated $\pi^\pm$);  background correlations of
decay photons. In this paper we consider all of these contributions
in detail and describe how to control for them in the
PHENIX experiment.

\section{Analysis}

This analysis is based on the data taken by PHENIX in Run3 (d+Au)
and Run4 (Au+Au). The total collected statistics is $\approx 3$
billion d+Au events and $\approx 900$ M Au+Au events. Details of
the PHENIX configuration in these runs can be found in references
\cite{PHENIX-dAu} and \cite{PHENIX-AuAu}, respectively.

\subsection{Apparatus effects}

Since correlation functions are rapidly rising functions at small
relative momenta any small distortion of the relative momentum for
real pairs, because of errors in reconstruction of close clusters
in the calorimeter ("cluster attraction") for example,  can lead
to the appearance of a fake bump in the correlation function.

To explore the influence of cluster interference in the calorimeter
EMCAL, we construct a set of correlation functions by applying
different cuts on the minimal distance between photon clusters in
EMCAL. To quantify the difference between these correlation
functions we fit them with a Gaussian and compare the extracted
correlation parameters. We find that for correlation functions
that include clusters with small relative distances there is strong
dependence on minimal distance cut, but for distance cuts above
24 cm (4-5 modules) the correlation parameters are independent of the
relative distance cut. This implies that with this distance cut the
apparatus effects are sufficiently small.

\subsection{Photon conversion, $\bar n$ annihilation, and similar  backgrounds}

The next class of possible backgrounds are processes in which one
real particle produces several clusters in the calorimeter close
to each other. These are processes like $\bar n$ annihilation in
the calorimeter producing several separated clusters, or photon
conversion in front of calorimeter, or residual correlations
between photons that belong to different $\pi^0$ in decays like
$\eta\to 3\pi^0\to6\gamma$. The common feature of this type of
process is that their strength is proportional to the number of
particles per event and not to the square of the number of
particles, as would be the case for Bose-Einstein correlations.

\begin{figure}[t]
\centerline{\psfig{file=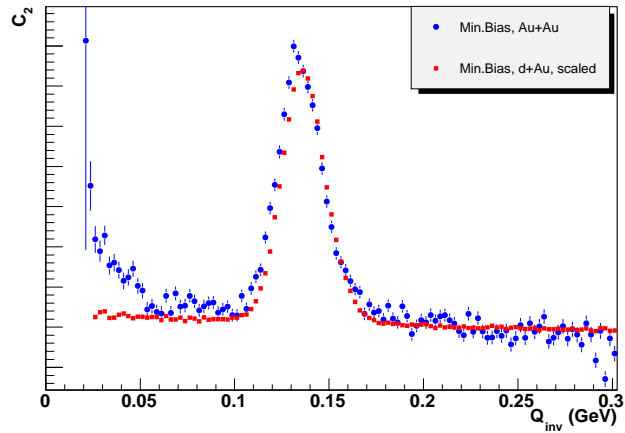,width=9.5cm}}
%\vspace*{8pt}
\caption{Two-photon correlation function measured in d+Au
collisions at $\sqrt{s_{NN}}=200$~GeV scaled to reproduce the
height of the $\pi^0$ peak in Au+Au collisions compared to the
same correlation function measured in Au+Au collisions at
$\sqrt{s_{NN}}=200$~GeV. Absolute vertical scale is omitted in
this technical plot.} \label{fig:dAu}
\end{figure}

To estimate the upper limit on these contributions, we compare
two-photon correlation functions, calculated in d+Au and Au+Au
collisions. For the moment we assume, that all correlations at
small relative momenta seen in d+Au collisions are due to the
background effects under consideration. Then we scale the
correlation function obtained in d+Au collisions with the number
of $\pi^0$ (that is we reproduce the height of the $\pi^0$ peak in
Au+Au collisions):
\begin{equation}\label{c2}
 C_2^{scaled}=1-\frac{h_\pi^{Au+Au}}{h_\pi^{d+Au}}(C_2-1).
\end{equation}

The result of this operation is shown in Fig.~\ref{fig:dAu}. We
find that the scaled d+Au correlation function lies well below
(close to unity) the correlation function calculated for Au+Au
collisions at small relative momenta.  From this we conclude that
the contribution from effects with strength proportional to the
first power of the number of particles is negligible in Au+Au
collisions.

\subsection{Charged and neutral hadron contamination}

Another possible source of distortion of the photon correlation function
is a contamination by (correlated) hadrons. Although we use rather
strict identification criteria for photons there still may be some
admixture of correlated hadrons contributing to the region of small
relative momenta.

\begin{figure}[th]
\centerline{\psfig{file=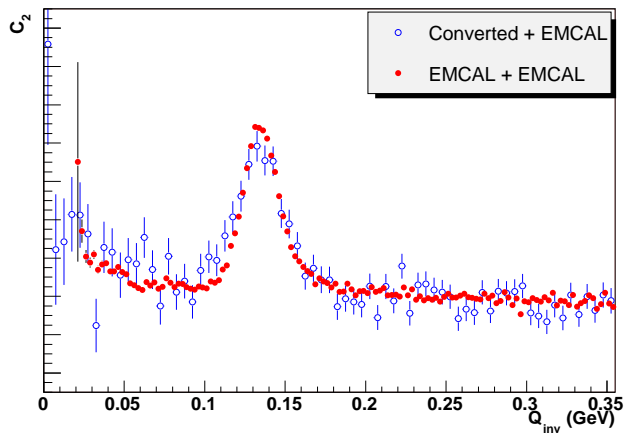,width=9.5cm}}
%\vspace*{8pt}
\caption{Comparison of two-photon correlation functions measured
in Au+Au collisions at $\sqrt{s_{NN}}=200$~GeV by two different
methods: both photons are registered in the EMCAL (closed) and one
photon is registered in EMCAL while the other is reconstructed
through its external conversion (open). Absolute vertical scale is
omitted in this technical plot.} \label{fig:conv}
\end{figure}

To exclude this possibility, we construct the two-photon
correlation function using one photon registered in the
calorimeter EMCAL and reconstructing the second photon from its
conversion into an $e^+e^-$ pair on the material of the beam pipe.
The photon sample, constructed using external conversions is
completely free from hadron contamination, so comparison of the
"standard" correlation function with the "pure" one allows to
estimate the contribution from non-photon contamination. This
comparison is shown in Fig.~\ref{fig:conv}. We find that the
correlation function constructed with the more pure photon sample
demonstrates a slightly larger correlation strength. This
demonstrates that the observed correlation is indeed a photon
correlation, while hadron contamination in the photon sample just
increases combinatorial background and reduces the correlation
strength. In addition, this comparison shows that we have properly
excluded the region of cluster interference. Due to deflection by
the magnetic field the electrons of the $e^+e^-$ conversion pair
hit the calorimeter far from the location of the pair photon used
in the correlation function and thus effects related to the
interference of close clusters are absent.

\subsection{Photon residual correlations}

The last possible source of the distortion of the photon
correlation function are residual correlations between photons. We  have already
demonstrated that the contributions of residual correlations between
photons in decays like $\eta\to3\pi^0\to 6\gamma$, with strength
proportional to $N_{part}$ and not $N_{part}^2$ is negligible
in Au+Au collisions. Below we consider other effects, which may
cause photon correlations. These are collective flow (and
jet-like correlations) and correlations between photons,
originated from decays of Bose-Einstein correlated mesons.
Collective (elliptic) flow as well as jet-like correlations are
long-range effects, resulting in correlations at relative angles much
larger than under consideration here (for example, the opening angle of a photon pair
with 20 MeV mass and $K_T = 500$~MeV is
% $\sim 8\times 10^{-2}$~rad
$\sim 5$ degrees). Monte-Carlo simulations demonstrate that flow and
jet-like contribution are indeed negligible.

\begin{figure}[th]
\centerline{\psfig{file=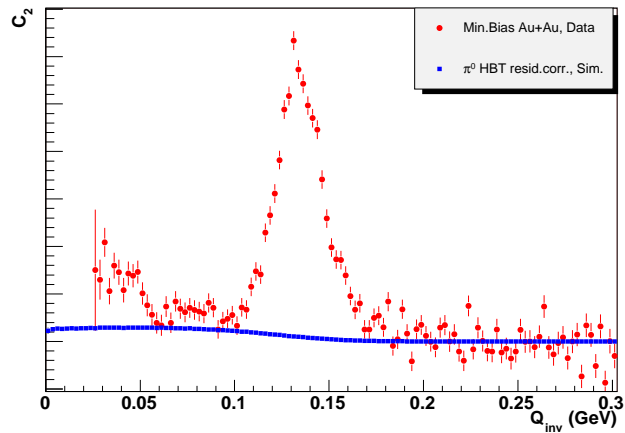,width=9.5cm}}
%\vspace*{8pt}
\caption{Comparison of two-photon correlation functions measured
in Au+Au collisions at $\sqrt{s_{NN}}=200$~GeV with Monte-Carlo
simulations of the contribution of residual correlations due to
decays of Bose-Einstein-correlated neutral pions. Absolute
vertical scale is omitted in this technical plot.}
\label{fig:pi0HBT}
\end{figure}

Potentially, the most serious distortion of the photon correlation
function are residual correlations between decay photons of
HBT-correlated $\pi^0$s. Monte-Carlo simulations show that this
contribution is not negligible, but has a rather specific shape
(see Fig.~\ref{fig:pi0HBT}), so that it does not distort the
photon correlation function at small $Q_{inv}$. This result can be
explained as follows. Let us consider two $\pi^0$s with zero
relative momentum. The distribution of decay photons is isotropic
in their rest frame, and the probability to find a collinear
photon pair ($Q_{inv}=0$) is suppressed due to phase space
reasons. The photon pair mass distribution has a maximum at
$2/3\,m_\pi$, not at zero. After convoluting with the pion
correlation function we find a step-like two-photon correlation
function\cite{Peressounko}. On the other hand, if one artificially
chooses photons with momentum along the direction of the parent
$\pi^0$ (e.g. by looking at photon pairs at very large $K_T$),
then the shape of the decay photon correlation function will
reproduce the shape of the parent $\pi^0$ correlation. This
probably explains the different shape of the residual correlations
due to decays of HBT-correlated $\pi^0$ found in\cite{STAR-ggHBT}.

\section{Conclusions}

We have presented the current status of analysis of direct photon
Bose-Einstein correlations in the PHENIX experiment. We are able
to measure the two-photon correlation function with a precision
sufficient to extract the direct photon correlations.
Correlation measurements in which one of the photon pair has converted
to an $e^+e^-$ pair have been used to provide an important cross-check.
We have demonstrated that all known backgrounds are
under control. The extraction of the correlation parameters of direct
photon pairs is in progress.

\end{document}